\documentclass[
twocolumn,
amsmath,amssymb,
aps,
]{revtex4-1}
\usepackage{graphicx}
\usepackage{hyperref}
\usepackage{dcolumn}
\usepackage{bm}
\usepackage{epstopdf}
\usepackage{romannum}
\usepackage{csquotes}
\usepackage[dvipsnames]{xcolor}

\begin{document}

\preprint{Sun/Anisotropy of FeTeSe}

\title{Comparative study of superconducting and normal-state anisotropy in Fe$_{1+y}$Te$_{0.6}$Se$_{0.4}$ superconductors with controlled amounts of interstitial excess Fe}

\author{Yue Sun$^1$}
\email{sunyue@phys.aoyama.ac.jp}
\author{Yongqiang Pan$^2$, Nan Zhou$^2$, Xiangzhuo Xing$^2$, Zhixiang Shi$^2$} 
\email{zxshi@seu.edu.cn}	
\author{Jinhua Wang$^3$, Zengwei Zhu$^3$, Akira Sugimoto$^4$, Toshikazu Ekino$^4$, Tsuyoshi Tamegai$^5$}
\author{Haruhisa Kitano$^1$}
\email{hkitano@phys.aoyama.ac.jp}

\affiliation{%
$^1$Department of Physics and Mathematics, Aoyama Gakuin University, Sagamihara 252-5258, Japan\\
$^2$School of Physics and Key Laboratory of MEMS of the Ministry of Education, Southeast University, Nanjing 211189, China\\
$^3$Wuhan National High Magnetic Field Center, School of Physics, Huazhong University of Science and Technology, Wuhan 430074, China\\
$^4$Graduate School of Advanced Science and Engineering, Hiroshima University, Higashi-Hiroshima 739-8521, Japan\\
$^5$Department of Applied Physics, The University of Tokyo, Tokyo 113-8656, Japan}

\date{\today}

\begin{abstract}
We report a systematic study of the superconducting (SC) and normal-state anisotropy of Fe$_{1+y}$Te$_{0.6}$Se$_{0.4}$ single crystals with controlled amounts of excess Fe ($y$ = 0, 0.07, and 0.14). The SC state anisotropy $\gamma_{H}$ was obtained by measuring the upper critical fields under high magnetic fields over 50 T for both $H\parallel ab$ and $H\parallel c$. On the other hand, the normal state anisotropy $\gamma_{\rho}$ was obtained by measuring the resistivity with current flowing in the $ab$ plane ($\rho_{ab}$) and along the $c$ axis ($\rho_c$). To precisely measure $\rho_{ab}$ and $\rho_c$ in the same part of a specimen avoiding the variation dependent on pieces or parts, we adopt a new method using a micro-fabricated bridge with an additional neck part along $c$ axis. The $\gamma_{H}$ decreases from a value dependent on the amount of excess Fe at $T_{\rm{c}}$ to a common value $\sim$ 1 at 2 K. The different $\gamma_{H}$ at $T_{\rm{c}}$ ($\sim$1.5 for $y$ = 0, and 2.5 for $y$ = 0.14) suggests that the anisotropy of effective mass $m_c^*/m_{ab}^*$ increases from $\sim$ 2.25 ($y$ = 0) to 6.25 ($y$ = 0.14) with the excess Fe. The almost isotropic $\gamma_{H}$ at low temperatures is due to the strong spin paramagnetic effect at $H\parallel ab$. By contrast, the $\gamma_{\rho}$ shows a much larger value of $\sim$ 17 ($y$ = 0) to $\sim$ 50 ($y$ = 0.14) at the temperature just above $T_{\rm{c}}$. Combined the results of $\gamma_{H}$ and $\gamma_{\rho}$ near $T_{\rm{c}}$, we found out that the discrepant anisotropies between the SC and normal states originates from a large anisotropy of scattering time $\tau_{ab}$/$\tau_c$ $\sim$ 7.8. The $\tau_{ab}$/$\tau_c$ is found to be independent of the excess Fe.  

\end{abstract}

\maketitle
\section{introduction}
Fe$_{1+y}$Te$_{1-x}$Se$_{x}$ compounds are unique in iron-based superconductors (IBSs) because of their structural simplicity, consisting of only FeTe/Se layers. They have attracted much interest both in the fundamental physics and application research. In the fundamental physics, the SC transition temperature ($T_{\rm{c}}$) is found to be remarkably enhanced by applying pressure \cite{MedvedevNatMat,GrestyJacs}, intercalating spacer layers \cite{BurrardNatMat,LiFeOHSeNatMater}, carrier doping by gating \cite{ShiogaiNatPhy,LeiPhysRevLett.116.077002}, and reducing the thickness to monolayer \cite{HeShaolongNatMat,GeNatMat}. A nematic state, which break the rotational symmetry, is observed in FeSe without long-range magnetic order \cite{WangNatMater2016,WenNatComm,WangNatComun2016}. The small Fermi energy, comparable to the superconducting gap size, indicates that superconductivity in Fe$_{1+y}$Te$_{1-x}$Se$_{x}$ may be in the crossover regime from Bardeen-Cooper-Schrieffer (BCS) to Bose-Einstein condensation (BEC) \cite{Kasahara18112014,LubashevskyNatPhy}. More interestingly, a topological surface superconductivity \cite{ZhangARPESScience,ZhangPengNatPhy} and the possible Majorana bound state have been observed \cite{WangMajoranaScience,MachidaNatMat}, which make Fe$_{1+y}$Te$_{1-x}$Se$_{x}$ the first high-temperature topological superconductor. In the view of application, the large upper critical field ($H_{\rm{c2}}$) and less toxic nature compared with iron pnictides make Fe$_{1+y}$Te$_{1-x}$Se$_x$ an ideal candidate for fabricating SC wires and tapes. In practice, the SC tapes with a large critical current density, over 10$^6$ A/cm$^2$ under self-field and over 10$^5$ A/cm$^2$ under 30 T at 4.2 K, have already been fabricated \cite{SiWeidongNatComm}. 

Determination of the anisotropy ($\gamma$) is crucial for both fundamental physics and practical applications \cite{GurevichRPP,Hosonoreview}. It provides information on the underlying electronic structure such as the Fermi-surface topology, and also shed light on the SC gap structure. In application, small $\gamma$ is advantageous for allowing high critical current density in the presence of magnetic field, due to the reduction of flux cutting effects and strong thermal fluctuations. Therefore, the anisotropy of Fe$_{1+y}$Te$_{1-x}$Se$_{x}$ is pivotal for both understanding the intriguing physics and the future application.   

In the SC state, $\gamma$ can be obtained by measuring $H_{\rm{c2}}$ or the penetration depth ($\lambda$). The former defines $\gamma_H$ = $H_{c2}^{ab}$/$H_{c2}^c$ = $\xi_{ab}$/$\xi_c$, where $\xi_{ab}$ and $\xi_c$ are the coherence lengths in the $ab$ plane and along $c$ axis, respectively. The later provides $\gamma_{\lambda}$ = $\lambda_c$/$\lambda_{ab}$, where $\lambda_c$ and $\lambda_{ab}$ are the penetration depths. Within the Ginzburg-Landau (GL) theory for a single-gap superconductor at the temperatures close to $T_{\rm{c}}$, $\gamma_H$ = $\sqrt{m_c^*/m_{ab}^*}$ = $\gamma_{\lambda}$, where $m_c^*$ and $m_{ab}^*$ are the effective masses along $c$ axis and in the $ab$ plane, respectively \cite{Tinkham,YuanAPL}. On the other hand, $\gamma$ in the normal state can be obtained by $\gamma_{\rho}$ = $\rho_c$/$\rho_{ab}$, where $\rho_c$ and $\rho_{ab}$ are the resistivity along the $c$ axis and in the $ab$ plane, respectively. In the approximation of isotropic scattering, $\rho_c$/$\rho_{ab}$ = $m_c^*/m_{ab}^*$. Therefore, $\gamma$ in the SC and normal state can be connected by the relation $\gamma_H$ ($\gamma_{\lambda}$) $\sim$ $\gamma_{\rho}^{1/2}$ \cite{TanatarPhysRevBAnisotropy122}.  

\begin{table}
	\caption{SC and normal-state anisotropies for typical IBSs. SC state anisotropy is calculated as $\gamma_H$ = $H_{c2}^{ab}$/$H_{c2}^c$, where $H_{c2}^{ab}$ and $H_{c2}^c$ are the upper critical fields in the $ab$ plane and along the $c$ axis. Normal-state anisotropy is obtained as $\gamma_{\rho}$ = $\rho_c$/$\rho_{ab}$, where $\rho_c$ and $\rho_{ab}$ are resistivity along $c$ axis and in the $ab$ plane. To compare with $\gamma_H$, $\gamma_{\rho}^{1/2}$ is calculated and presented in the table. Anisotropies for Fe$_{1.0}$Te$_{0.6}$Se$_{0.4}$, Fe$_{1.07}$Te$_{0.6}$Se$_{0.4}$, and Fe$_{1.14}$Te$_{0.6}$Se$_{0.4}$ are the results of the current research.}
	\begin{ruledtabular}
		\begin{tabular}{c c c c }
			& $\gamma_H$ (SC state) & $\gamma_{\rho}^{1/2}$ (normal state)    \\	
			Ba(Fe$_{1-x}$Co$_x$)$_2$As$_2$     & 1.5$\sim$2.0 \cite{YamamotoAPL}  & 1.4$\sim$2.1\cite{TanatarPhysRevBAnisotropy122}        \\
			Ba$_{1-x}$K$_x$Fe$_2$As$_2$     & 1$\sim$2 \cite{Yuannature}  & 3.2$\sim$5.5 \cite{ZverevJETP}         \\
			BaFe$_2$(As$_{1-x}$P$_x$)$_2$     & 1.5$\sim$2.6 \cite{ChaparroBaFe2AsPHc2,Miuranatcomm}  & 2$\sim$2.8 \cite{TanatarPhysRevB.87.104506}       \\
			LiFeAs   & 1.5$\sim$2.5 \cite{ZhangLiFeAsPhysRevB.83.174506}  & 1.2$\sim$1.9 \cite{SongLiFeAs_gammaAPL}       \\
			KFe$_2$As$_2$    & 3.5$\sim$5.5 \cite{TerashimaKFe2As2_gamma}  & 3.2$\sim$6.3 \cite{TerashimaKFe2As2_gamma}     \\
			SmFeAsO$_{1-x}$F$_x$    & 4$\sim$5 \cite{MollSmFeAsOFNatMat}  & 1.4$\sim$3.5 \cite{MollSmFeAsOFNatMat}      \\
			Ca$_{1-x}$La$_x$FeAs$_2$ & 4.9-5.2 \cite{XingNJP}  & 3.9-5.5 \cite{JiangPhysRevB.93.054522}   \\
			Fe$_{1.0}$Te$_{0.6}$Se$_{0.4}$  & 1$\sim$1.6  & 2.5$\sim$4       \\
			Fe$_{1.07}$Te$_{0.6}$Se$_{0.4}$  & 1$\sim$1.8  & -       \\
			Fe$_{1.14}$Te$_{0.6}$Se$_{0.4}$   & 1$\sim$2.5  & 5.7$\sim$7   \\     
			Fe$_{1.18}$Te$_{0.6}$Se$_{0.4}$   & -  & 7.7$\sim$8.8 \cite{LiuPhysRevB.80.174509}
			
		\end{tabular}
	\end{ruledtabular}
\end{table}

So far, the relation $\gamma_H$ $\sim$ $\gamma_{\rho}^{1/2}$ has already been verified in most IBSs, as summarized in Table \Romannum{1}. However, the relation seems to be violated in Fe$_{1+y}$Te$_{1-x}$Se$_{x}$. A small SC-state anisotropy, $\gamma_H$($\gamma_{\lambda}$) $<$ 3, has already been confirmed by several previous reports \cite{FangMinghuPhysRevB.81.020509,KhimPhysRevB.81.184511,BendelePhysRevB_lambda}. By contrast, an unexpectedly large normal-state anisotropy $\gamma_{\rho}$ $\sim$ 50-70 was reported \cite{NojiFeTeSeannealing}. Such a discrepancy between the SC and normal-state anisotropies still remains unresolved, which confuses both the study of fundamental physics and the application of Fe$_{1+y}$Te$_{1-x}$Se$_{x}$.

In this report, we successfully resolved the discrepancy by systematically probing the SC and normal-state anisotropies of Fe$_{1+y}$Te$_{0.6}$Se$_{0.4}$ single crystals with different amounts of excess Fe. Such discrepancy is demonstrated to originate from a large anisotropy in scattering times $\tau_{ab}$/$\tau_c$ $\sim$ 7.8 in the normal state.

\section{experiment}
Fe$_{1+y}$Te$_{0.6}$Se$_{0.4}$ single crystals were grown by the self-flux method as described in detail elsewhere \cite{SunSUSTFeSeannealing}. The as-grown crystals usually contain some amounts (represented by $y$) of excess Fe residing in the interstitial sites of the Te/Se layer. The excess Fe can be removed and its amount can be tuned by post annealing \cite{SunSciRep,SunJPSJTeannealing,SunJPSJSeSanneal,Sun_sustreview}. After annealing, a series of single crystals with different amounts of excess Fe can be prepared. More details about the crystal preparation, excess Fe, and the basic properties can be found in our recent review paper \cite{Sun_sustreview}. The inductively-coupled plasma (ICP) atomic emission spectroscopy and the scanning tunneling microscopy (STM) were used for detecting the amount of excess Fe. STM images were obtained by a modified Omicron LT–UHV–STM system \cite{SugimotoJPSJ}. The sample was cleaved in situ at 4 K in an ultra-high vacuum chamber of $\sim$ 10$^{-8}$ Pa to obtain fresh and unaffected crystal surface. Resistivity measurements were performed by the four-probe method. The electrical transport measurements under high magnetic field were performed at Wuhan High Magnetic Field Center, China. The bridges in the $ab$ plane and along the $c$ axis used for the measurements of normal state anisotropy, as shown schematically in the insets of Figs. 3(a) and 3(b), were fabricated by using the focused ion beam (FIB) technique \cite{SunPhysRevB.101.134516,KakehiIEEE,Kakizaki_JJAP}.

\section{results and discussion}

\begin{figure}\center
	\includegraphics[width=8.5cm]{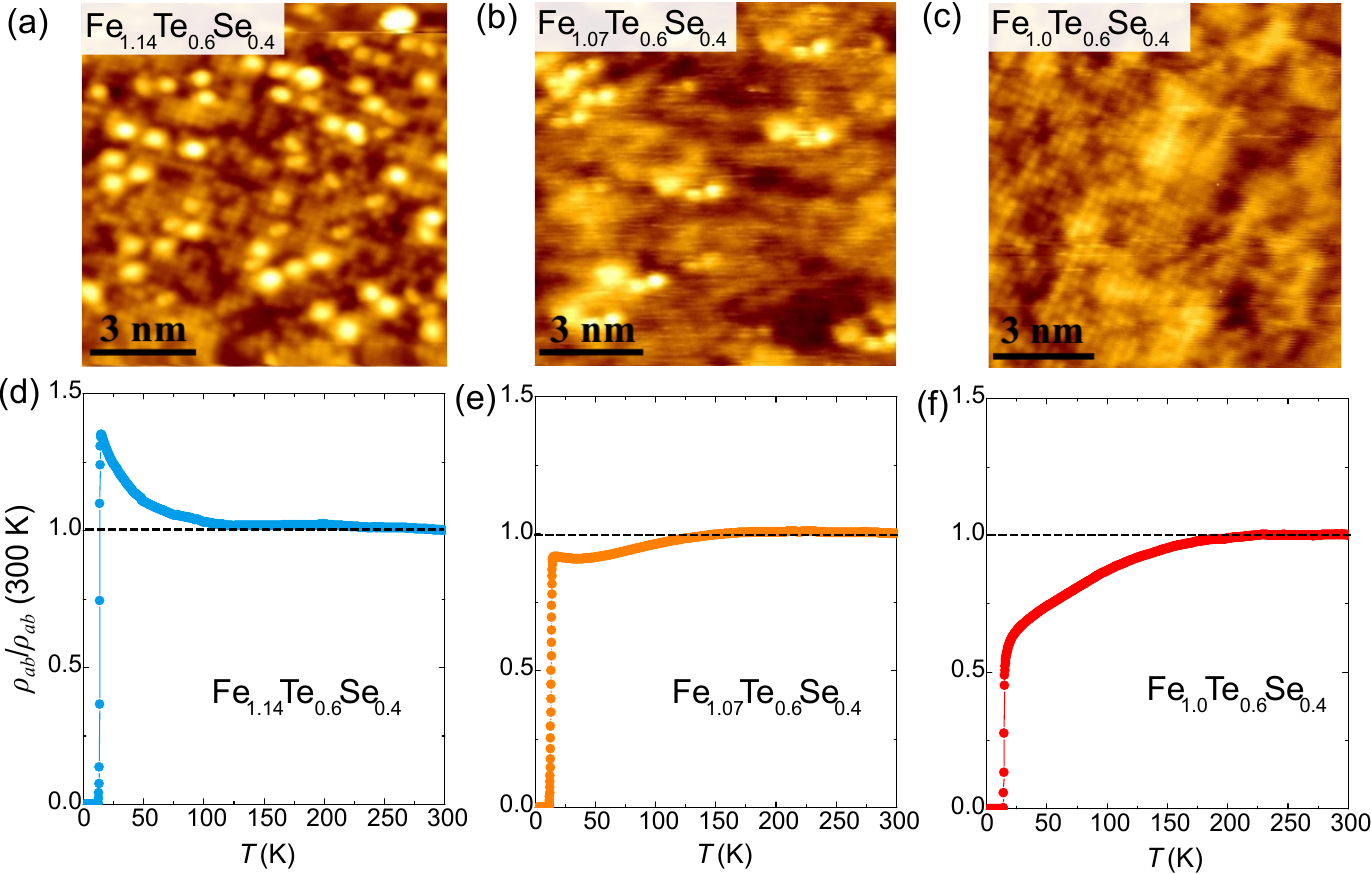}
	\caption{STM images for (a) Fe$_{1.14}$Te$_{0.6}$Se$_{0.4}$, (b) Fe$_{1.07}$Te$_{0.6}$Se$_{0.4}$, and (c) Fe$_{1.0}$Te$_{0.6}$Se$_{0.4}$ single crystals. The bright spots in (a) and (b) correspond to the excess Fe, which disappear in (c). (a) and (c) have been used in our previous publication \cite{SunSciRep}. Temperature dependence of the in-plane resistivities scaled by the values at 300 K for (d) Fe$_{1.14}$Te$_{0.6}$Se$_{0.4}$, (e) Fe$_{1.07}$Te$_{0.6}$Se$_{0.4}$, and (f) Fe$_{1.0}$Te$_{0.6}$Se$_{0.4}$.}\label{}
\end{figure}

In our as-grown single crystals, the amount of excess Fe is $\sim$ 14\% as analyzed by ICP atomic emission spectroscopy. Although the excess Fe may be removed after annealing, it should still remain in the crystal, mainly on the surface, in some form of oxides \cite{Sun_sustreview}. Therefore, traditional compositional analysis methods such as the ICP, energy dispersive X-ray spectroscopy (EDX) and electron probe microanalyzer (EPMA) cannot precisely detect the amount change of excess Fe. To precisely determine the change in the number of excess Fe, we employ the STM measurement, which has atomic resolution. The excess Fe occupies the interstitial site in the Te/Se layer, and the previous report proved that the cleaved Fe$_{1+y}$Te$_{1-x}$Se$_x$ single crystal possesses only the termination layer of Te/Se, which guarantee that the STM can directly observe the excess Fe in Te/Se layer without the influence of neighboring Fe layers \cite{MasseePhysRevB.80.140507}. Fig. 1(a) shows the STM image for the as-grown crystal. There are several randomly distributed bright spots in the image, which represent the excess Fe according to the previous STM analysis \cite{MasseePhysRevB.80.140507,HanaguriScience,UKITAPhysicaC}. After annealing, the amount of bright spots, i.e. the excess Fe, is obviously reduced as shown in Fig. 1(b), and disappears in Fig. 1(c). By counting the number of the bright spots in the STM images together with the ICP result of 14\% excess Fe for the as-grown crystal, the amount of excess Fe in Figs. 1(b) and 1(c) can be estimated as $\sim$ 7\% and 0, respectively. Hence, the three crystals are labeled as Fe$_{1.14}$Te$_{0.6}$Se$_{0.4}$, Fe$_{1.07}$Te$_{0.6}$Se$_{0.4}$, and Fe$_{1.0}$Te$_{0.6}$Se$_{0.4}$ in the rest of this article.

Figs. 1(d)-1(f) show the temperature dependence of resistivities for the three crystals, scaled by the values at 300 K. All the crystals manifest a similar onset of $T_{\rm{c}}$ $\sim$ 15 K. However, the temperature dependent behaviors for the resistivity are quite different. Fe$_{1.14}$Te$_{0.6}$Se$_{0.4}$ manifests a semiconducting behavior (d$\rho$/d$T$ $<$ 0) when the temperature approaches to $T_{\rm{c}}$. The residual resistivity ratio RRR, defined as $\rho$(300 K)/$\rho$($T_{\rm{c}}^{\rm{onset}}$), is estimated as $\sim$ 0.74. For Fe$_{1.07}$Te$_{0.6}$Se$_{0.4}$, the semiconducting behavior is suppressed, and replaced by a temperature-independent behavior with RRR = 0.92. On the other hand, resistivity for Fe$_{1.0}$Te$_{0.6}$Se$_{0.4}$ manifests a metallic behavior (d$\rho$/d$T$ $>$ 0) with RRR = 2. These observations suggest that the semiconducting behavior (d$\rho$/d$T$ $<$ 0) in Fe$_{1.14}$Te$_{0.6}$Se$_{0.4}$ originates from the localization effect of excess Fe, which can be suppressed by removing the excess Fe \cite{LiuPhysRevB.80.174509,SunPRB2014}. More details about the transport properties such as the Hall effect and magnetoresistance have been reported in our previous publications \cite{SunPRB2014,SunSciRep2016}.               

\begin{figure*}\center
	\includegraphics[width=17cm]{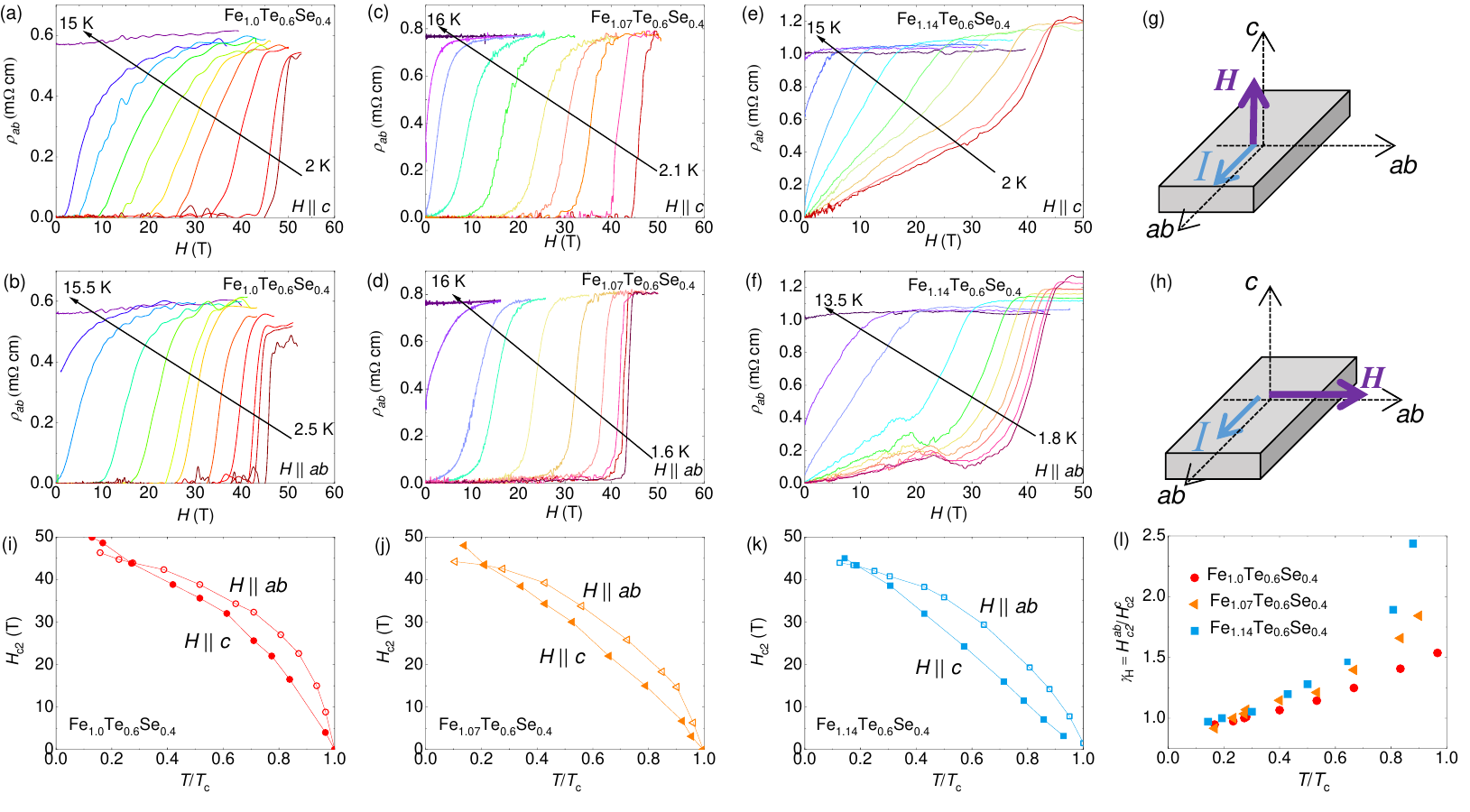}
	\caption{The magnetic field dependence of the in-plane resistivity $\rho_{ab}$ for Fe$_{1.0}$Te$_{0.6}$Se$_{0.4}$ with (a) $H\parallel c$ at 2, 2.6, 4.2, 6, 8, 10, 11, 12, 13, and 15 K, (b) $H\parallel ab$ at 2.5, 3.5, 4.2, 5, 8, 10, 11, 12, 13, 14, 14.5 and 15 K, for Fe$_{1.07}$Te$_{0.6}$Se$_{0.4}$ with (c) $H\parallel c$ at 2.1, 3.2, 4.2, 6, 7.5, 9, 10.5, 13.5, 14, and 16 K, (d) $H\parallel ab$ at 1.6, 3.2, 4.2, 6, 8, 11, 12.5, 13.2, 14, and 16 K, for Fe$_{1.14}$Te$_{0.6}$Se$_{0.4}$ with (e) $H\parallel c$ at 2, 2.6, 4.3, 6, 8, 10, 11, 12, 13, and 15 K, (f) $H\parallel ab$ at 1.8, 2.5, 3.5, 4.2, 5.5, 6.5, 9, 11.5, 12.5, and 13.5 K. Schematics of the experimental configuration for the resistivity measurements with (g) $H\parallel c$ and (h) $H\parallel ab$, respectively.  Reduced temperature ($T/T_c$) dependence of upper critical fields for (i) Fe$_{1.0}$Te$_{0.6}$Se$_{0.4}$, (j) Fe$_{1.07}$Te$_{0.6}$Se$_{0.4}$, and (k) Fe$_{1.14}$Te$_{0.6}$Se$_{0.4}$, where the solid and open symbols represent $H\parallel c$ and $H\parallel ab$, respectively. (l) Temperature dependence of the anisotropies in the SC state for the three crystals.}\label{}
\end{figure*}

To probe the anisotropy in the SC state, the SC transition was measured under a high magnetic field over 50 T for the three crystals. Figs. 2(a)-2(f) show the in-plane resisivity $\rho_{ab}$ of the three crystals as a function of the magnetic field along the $c$ axis ($H\parallel c$) and parallel to the $ab$ plane ($H\parallel ab$). $H_{c2}^{ab}$ (open symbols) and $H_{c2}^c$ (solid symbols) for the three crystals are determined by the 90\% of the resistivity value just above the	SC transition. (Due to the broad transition under $H\parallel c$ for Fe$_{1.14}$Te$_{0.6}$Se$_{0.4}$, the criteria of 50\% and 10\% of the resistivity value cannot be obtained at high temperatures.) Clearly, with such large field, we can reach the $H_{c2}$ down to very low temperatures $\sim$ 2 K, which is over 44 T for both $H\parallel ab$ and $H\parallel c$ in all the three crystals (see Figs. 2(i) - 2(k)). The obtained $H_{c2}$ is larger than the expected Pauli-limiting field estimated as $H_p$(0) = 1.86 $T_{\rm{c}}$ $\sim$ 27 T for a weak-coupling BCS superconductor, which indicates that the spin paramagnetic effect plays an important role in the determination of $H_{c2}$(0). On the other hand, $H_{c2}^{ab}$ shows a convex shape with similar curvatures for all the three crystals. The convex shape in $H_{c2}^{ab}$ is a common feature for IBSs \cite{FangMinghuPhysRevB.81.020509,KhimPhysRevB.81.184511,Yuannature,ZhangLiFeAsPhysRevB.83.174506,FuchsNJP}, which is usually explained by the strong spin paramagnetic effect with relative large Maki parameter $\alpha$ within the Werthamer-Helfand-Hohenberg (WHH) theory \cite{WHHmodel}. Therefore, almost the same behavior of $H_{c2}^{ab}$ for the three crystals indicates that the excess Fe has little effect on the spin paramagnetic effect for $H \parallel ab$. 

On the other hand, $H_{c2}^c$ for most IBSs manifests a nearly linear behavior (or less convex than $H_{c2}^{ab}$), suggesting that the spin paramagnetic effect for $H \parallel c$ is negligible (or much smaller than $H \parallel ab$) \cite{FangMinghuPhysRevB.81.020509,KhimPhysRevB.81.184511,Yuannature,ZhangLiFeAsPhysRevB.83.174506}. For Fe$_{1+y}$Te$_{1-x}$Se$_x$, both the convex and linear $H_{c2}^c$ have been reported previously \cite{FangMinghuPhysRevB.81.020509,KhimPhysRevB.81.184511}. In our case, $H_{c2}^c$ for Fe$_{1.14}$Te$_{0.6}$Se$_{0.4}$ shows a linear behavior, while a slightly convex behavior (with smaller curvature than $H_{c2}^{ab}$) is observed in Fe$_{1.07}$Te$_{0.6}$Se$_{0.4}$ and Fe$_{1.0}$Te$_{0.6}$Se$_{0.4}$. Our results reveal that the previous controversy in the $H_{c2}^c$ of Fe$_{1+y}$Te$_{1-x}$Se$_x$ is due to the sample dependence of excess Fe. For the crystals free from or with small amount of excess Fe, the spin paramagnetic effect is finite for $H \parallel c$, although smaller than that for $H \parallel ab$. However, the spin paramagnetic effect for $H \parallel c$ is is more easily suppressed by excess Fe, and becomes almost negligible in crystal with too much excess Fe. We also note that $H_{c2}^c$ for Fe$_{1.14}$Te$_{0.6}$Se$_{0.4}$ shows a weaker rise close to $T_{\rm{c}}$ than those for Fe$_{1.0}$Te$_{0.6}$Se$_{0.4}$ and Fe$_{1.07}$Te$_{0.6}$Se$_{0.4}$, suggesting a more strongly-divergent behavior of $\xi_{ab}$ close to $T_{\rm{c}}$. This leads to the finite difference of the SC state anisotropy dependent on the excess Fe, as will be discussed below.  

Due to the convex shape, $H_{c2}^{ab}$ finally meets $H_{c2}^c$ at low temperatures for all the three crystals, which means that the $H_{c2}$ becomes isotropic. With further decreasing temperature, $H_{c2}^{ab}$ becomes even smaller than $H_{c2}^c$. Such a crossover behavior is a unique feature of Fe$_{1+y}$Te$_{1-x}$Se$_x$, which is not observed in other IBSs \cite{GurevichRPP}. In a similar compound FeSe, a high-filed phase was observed at low temperatures, and suggested to originate from the Fulde-Ferrel-Larkin-Ovchinnikov (FFLO) state \cite{KasaharaPhysRevLettFeSeFFLO}. The large value of Maki parameter $\alpha$ and the possibility of FFLO state in FeTe$_{1-x}$Se$_x$ have also been discussed previously \cite{KhimPhysRevB.81.184511}. However, the realization of the FFLO state usually needs the crystal to be in the clean limit, i.e. the mean free path ($\ell$) should be much larger that the coherence length ($\xi$). According to the expressions \textit{$\ell$} = $\frac{\pi{c}\hbar}{{N}{e^2}{k_F}{\rho_0}}$ \cite{KasaharaPhysRevLettFeSeFFLO}, where \textit{c} is the lattice parameter, \textit{N} is the number of formula units per unit cell, \textit{$k$}$_{\rm{F}}$ $\sim$ 1.0 nm$^{-1}$ \cite{LubashevskyNatPhy} is the Fermi wave vector, and \textit{$\rho$}$_{0}$ $\sim$ 200 $\mu\Omega$ \cite{SunAPEX} is the residual resistivity, \textit{$\ell$} for Fe$_{1.0}$Te$_{0.6}$Se$_{0.4}$ is estimated as $\sim$ 1.8 nm. \textit{$\ell$} is smaller than $\xi$ $\sim$ 2.8 nm \cite{LeiHechangPhysRevcoherencelenth}, implying that the crystal is in the dirty limit rather than the clean limit. On the other hand, considering the fact that the transition from BCS state to the FFLO state is of first order, the FFLO state should be readily destroyed by disorders. It is obviously in stark contrast to our observations that the crossover of $H_{c2}^{ab}$ and $H_{c2}^c$ is almost identical in the three crystals containing different amounts of excess Fe. Therefore, the above discussion has ruled out the possibility of the FFLO state in Fe$_{1+y}$Te$_{0.6}$Se$_{0.4}$. A possible origin of the crossover behavior in the $H_{c2}$ is the multi-band effect. The upturn in $H_{c2}^c$ may be due to the contribution from another band. Similar upturn behavior has also been observed in S-doped FeSe \cite{AbdelHafiezPhysRevB.91.165109} and Ba$_2$Ti$_2$Fe$_2$As$_4$O \cite{Abdel-HafiezPhysRevB.97.115152}. We want to point out that such upturn behavior only occurs at low temperatures, which will not affect the value of anisotropy at high temperatures close to $T_{\rm{c}}$.      

The SC state anisotropies for the three crystals estimated as $\gamma_H$ = $H_{c2}^{ab}$/$H_{c2}^c$ are shown in Fig. 2(l). At low temperatures, $\gamma_H$ becomes isotropic for all the three crystals. Then, $\gamma_H$ gradually increases with increasing temperature, and manifests a stronger increase with excess Fe. The temperature dependence of $\gamma_H$ has been discussed by using $\Delta(k_z)$=$\Delta_0$(1+$\eta$cos$k_z\alpha$), including the coefficient $\eta$ for the $k_z$ dispersion of the gap \cite{Kogan_anisotropy_review}. $\gamma_H$ reaches a value $\sim$ 1.5 close to $T_{\rm{c}}$ for Fe$_{1.0}$Te$_{0.6}$Se$_{0.4}$. On the other hand, $\gamma_H$ close to $T_{\rm{c}}$ slightly increases with increasing the amount of excess Fe, and reaches a value $\sim$ 2.5 for Fe$_{1.14}$Te$_{0.6}$Se$_{0.4}$. Within the anisotropic three-dimensional GL-theory for a single-gap superconductor, $\gamma_H$ = $H_{c2}^{ab}$/$H_{c2}^c$ = $\sqrt{m_c^*/m_{ab}^*}$, through the anisotropy of the GL coherence lengths. The anisotropy of effective mass $m_c^*/m_{ab}^*$ is estimated as 2.25, 3.24, and 6.25 for Fe$_{1.0}$Te$_{0.6}$Se$_{0.4}$, Fe$_{1.07}$Te$_{0.6}$Se$_{0.4}$, and Fe$_{1.14}$Te$_{0.6}$Se$_{0.4}$, respectively. The influence of excess Fe on the $m_{ab}^*$ of the heavy band has been reported in the previous ARPES measurements \cite{ShaharARPESSciAdv}. Here, our results reveal that the anisotropy of $m_c^*/m_{ab}^*$ is also strongly affected by the excess Fe.  
                                                                   
\begin{figure}\center
	\includegraphics[width=8.5cm]{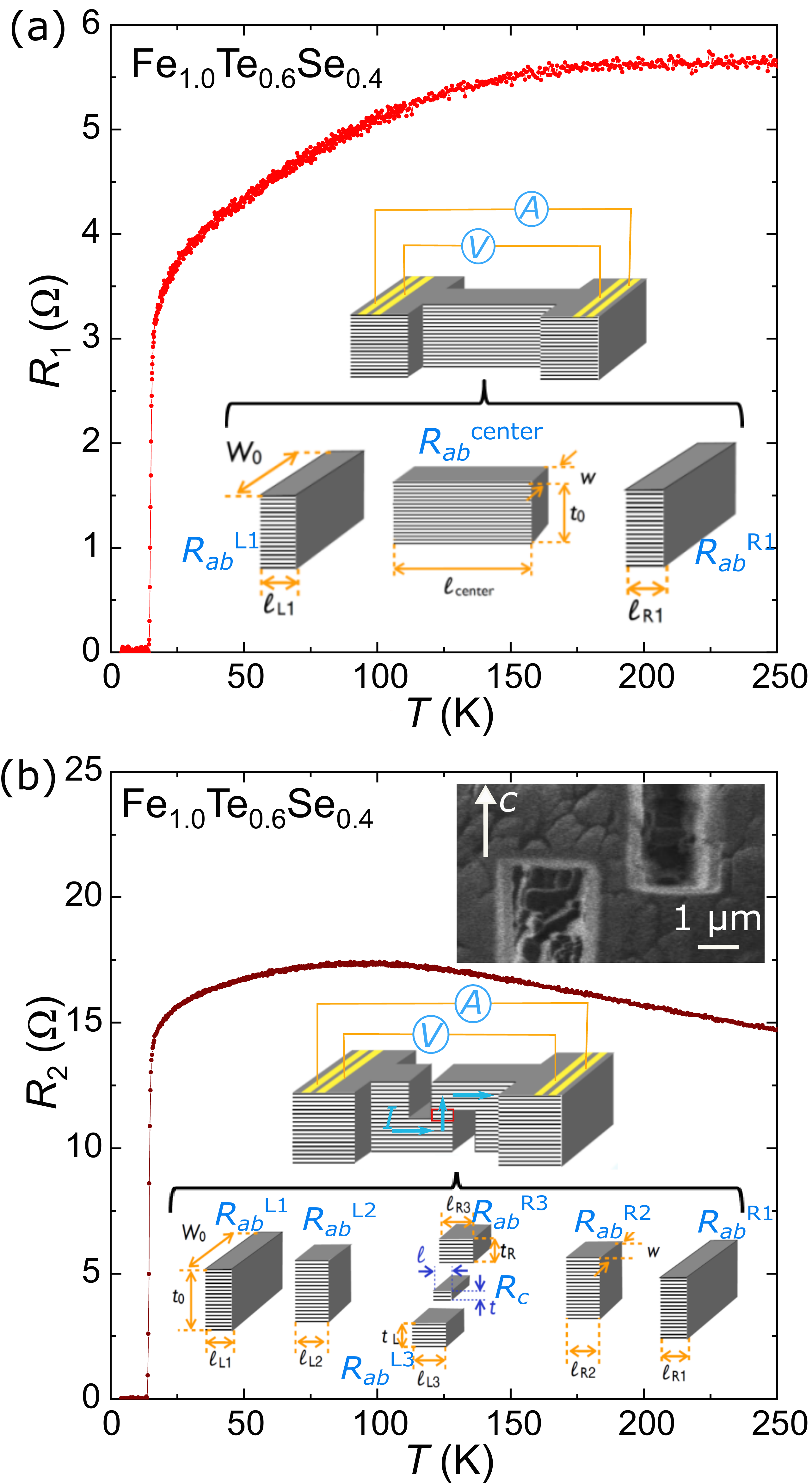}
	\caption{(a) Temperature dependence of the resistivity for the in-plane bridge $R_1$. Insets show the sketch and corresponding series resistors for $R_1$. (b) Temperature dependence of the resistivity for the further fabricated structure with two overlapped slits along the $c$ axis. Upper inset is the scanning ion microscopy image of the structure. Lower inset shows the sketch and corresponding series resistors for $R_2$.}\label{}
\end{figure}

In order to estimate the normal-state anisotropy, we need to measure the resistivity both in the $ab$ plane ($\rho_{ab}$) and along the $c$ axis ($\rho_c$). To measure $\rho_c$ for bulk sample, the specific configuration of contact electrodes is required for layered superconductors such as IBSs. This leads to a problem that $\rho_c$ and $\rho_{ab}$ are obtained from different samples. Here, we report a method to obtain the $\rho_c$ measured in a part of the region where $\rho_{ab}$ was measured, by using a $c$-axis neck structure fabricated additionally in the in-plane bridge. To fabricate the $c$-axis bridge, the crystal was first cleaved into a slice with $\sim$10 $\mu$m in thickness, by using scotch tape. The slice was glued on a sapphire substrate, and sputtered by four Au contacts to improve the electric contact. Then the sliced crystal was etched by using FIB and a narrow in-plane bridge with a width of $\sim$1 $\mu$m was fabricated between voltage terminals, as shown schematically in the inset of Fig. 3(a). The resistance $R_1$ for the in-plane bridge is measured by four probe method, which can be treated as a sum of three resistances in series, and expressed as 
\begin{equation}
\label{eq.1}
\begin{split}
R_1&=R_{ab}^{L1}+R_{ab}^{center}+R_{ab}^{R1}\\
&=\rho_{ab}(\frac{l_{L1}}{t_0W_0}+\frac{l_{center}}{t_0w}+\frac{l_{R1}}{t_0W_0}),
\end{split}
\end{equation}
where $t_0$ is the thickness, $w$ and $W_0$ are the width, $l_{L1}$, $l_{center}$, and $l_{R1}$ are the length of the three parts, as shown in the inset of Fig. 3(a). Temperature dependence of $R_1$ is shown in the main panel of Fig. 3(a).    

After measuring of $R_1$, two separated slits were further fabricated in the sidewalls of the in-plane bridge to make a small neck along the $c$ axis, as shown in the upper inset of Fig. 3(b). The length of $c$ axis neck was adjusted by a vertical overlap between the two slits (typically $\sim$1 $\mu$m). Such a crank structure enforces the current to flow along the $c$ axis in the bridge region as marked by the rectangular frame in the lower inset of Fig. 3(b). The whole resistance $R_2$ for this device can be treated as a sum of seven resistances in series as shown schematically in the lower inset of Fig. 3(b). The current flows along the $ab$ plane in the left and right three parts, while it flows along the $c$ axis in the center one with dimension of $l\times w \times t$. Therefore, $R_2$ can be expressed as
\begin{equation}
\label{eq.2}
\begin{split}
R_2&=R_{ab}^{L1}+R_{ab}^{L2}+R_{ab}^{L3}+R_{c}+R_{ab}^{R3}+R_{ab}^{R2}+R_{ab}^{R1}\\
&=\rho_{ab}(\frac{l_{L1}}{t_0W_0}+\frac{l_{L2}}{t_0w}+\frac{l_{L3}}{t_Lw})+\rho_c(\frac{t}{lw})\\
&+\rho_{ab}(\frac{l_{R3}}{t_Rw}+\frac{l_{R2}}{t_0w}+\frac{l_{R1}}{t_0W_0}).
\end{split}
\end{equation}
$\rho_{ab}$ and $\rho_c$ can be simply estimated by solving Eqs. (1) and (2). By this method, $\rho_{ab}$ and $\rho_c$ are obtained from almost the same region in an identical crystal, therefore they are not affected by the sample-dependent variations.  

\begin{figure}\center
	\includegraphics[width=8.5cm]{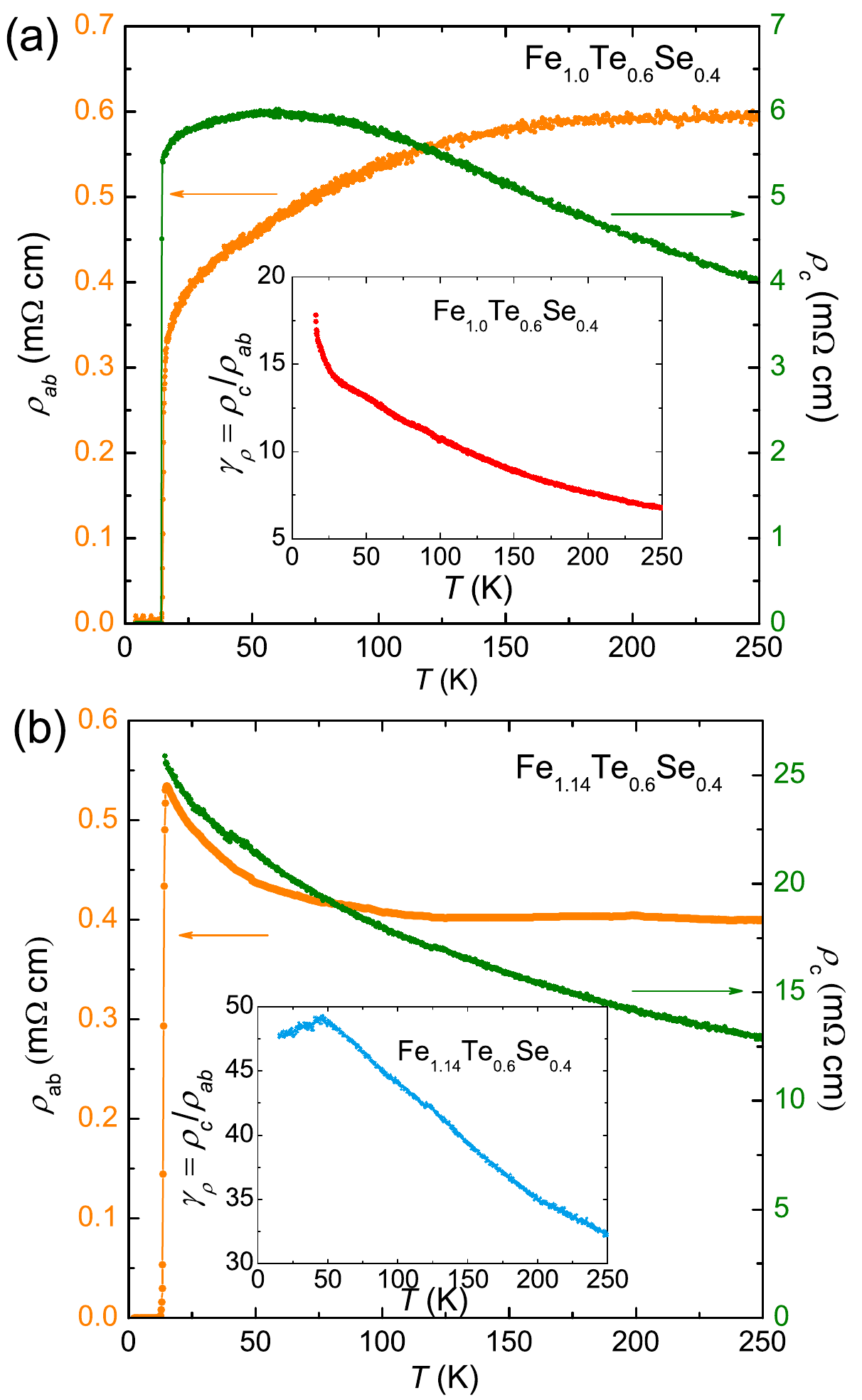}
	\caption{Temperature dependence of the in-plane (left axis) and out-of-plane (right axis) resistivity $\rho_{ab}$ and $\rho_c$ for the (a) Fe$_{1.0}$Te$_{0.6}$Se$_{0.4}$ and (b) Fe$_{1.14}$Te$_{0.6}$Se$_{0.4}$. Insets plot the normal-state anisotropy $\gamma_{\rho}$ = $\rho_c$/$\rho_{ab}$ as a function of temperature.}\label{}
\end{figure}

$\rho_{ab}$ and $\rho_c$ for the Fe$_{1.0}$Te$_{0.6}$Se$_{0.4}$ obtained by the above method are shown in the main panel of Fig. 4(a). Temperature dependence of $\rho_{ab}$ shows similar behavior as the bulk one [see Fig. 1(f)], which confirms that FIB fabrication will not introduce visible damage in the bridge part. In contrast to the $\rho_{ab}$, $\rho_c$ increases slightly with decreasing temperature down to 60 K, then it shows a metallic behavior down to $T_{\rm{c}}$. Similar temperature dependent behavior of $\rho_c$ was also reported previously by using the conventional method for bulk samples \cite{LiuPhysRevB.80.174509}. Normal state anisotropy $\gamma_{\rho}$ calculated as $\rho_c$/$\rho_{ab}$ for Fe$_{1.0}$Te$_{0.6}$Se$_{0.4}$ is shown in the inset of Fig. 4(a). $\gamma_{\rho}$ is $\sim$ 7 at 250 K, and gradually increases with decreasing temperature. Below $\sim$ 30 K, the increment accelerates, and $\gamma_{\rho}$ finally reaches a value of $\sim$ 17 just above $T_{\rm{c}}$. On the other hand, $\rho_c$ for Fe$_{1.14}$Te$_{0.6}$Se$_{0.4}$ with abundant excess Fe continues to increase with cooling down in the whole temperature range [see Fig. 4(b)]. Besides, the SC transition is not observed in $\rho_c$, which is due to the fact that superconductivity in Fe$_{1.14}$Te$_{0.6}$Se$_{0.4}$ is filamentary as proved by the absent of SC transition in bulk measurements such as specific heat and magnetization \cite{SunSciRep}. Our observation indicates that the filamentary superconductivity in crystals with abundant excess Fe is localized, andmay not show up in a small region where we probe $\rho_c$. The $\gamma_{\rho}$ for Fe$_{1.14}$Te$_{0.6}$Se$_{0.4}$ increases with decreasing temperature, while it decreases slightly below $\sim$ 50 K [see the inset of Fig. 4(b)]. In contrast to Fe$_{1.0}$Te$_{0.6}$Se$_{0.4}$, $\gamma_{\rho}$ for Fe$_{1.14}$Te$_{0.6}$Se$_{0.4}$ manifests a much larger value ranging from 32 - 50. Such larger normal-state anisotropy is close to that reported previously \cite{NojiFeTeSeannealing}. Our results reveal that the normal-state anisotropy is strongly affected by the amount of excess Fe.

To directly observe the temperature evolution of anisotropy in the whole temperature range from the SC to normal state, we summarized the results of $\gamma_H$ and $\gamma_{\rho}$ in Fig. 5. For comparison, the anisotropy $\gamma_{\lambda}$ estimated from penetration depth measurements (the amount of excess Fe was claimed to be $\sim$ 0) \cite{BendelePhysRevB_lambda}, and the $\gamma_{\rho}$ calculated from crystal with more excess Fe ($y$ = 0.18) \cite{LiuPhysRevB.80.174509} are also included. The normal-state anisotropy is compared with $\gamma_{H}$ by using a square root of $\gamma_{\rho}$, since $\gamma_{H}$ $\sim$ $\gamma_{\rho}^{1/2}$ is expected in the isotropic scattering case. In the SC state, both $\gamma_{H}$ and $\gamma_{\lambda}$ show relatively small values $<$3. On the other hand, $\gamma_{\lambda}$ increases with decreasing temperature, while $\gamma_H$ decreases with decreasing temperature. The different temperature dependence of $\gamma_{\lambda}$ and $\gamma_H$ in FeTe$_{1-x}$Se$_{x}$ has already been discussed in the previous report \cite{BendelePhysRevB_lambda}, and was also observed in other IBSs \cite{Prozorov_Review_Penetration} and MgB$_2$ \cite{MgB2anisotropy_PhysRevLett.95.097005}. It may originate from the multiband effect, where the contributions of electronic bands with different $k$-dependent Fermi velocities and gap values lead to different ratios of $\gamma_{\lambda}$ and $\gamma_H$ \cite{KonczykowskiPhysRevB.84.180514}.

\begin{figure}\center
	\includegraphics[width=8.5cm]{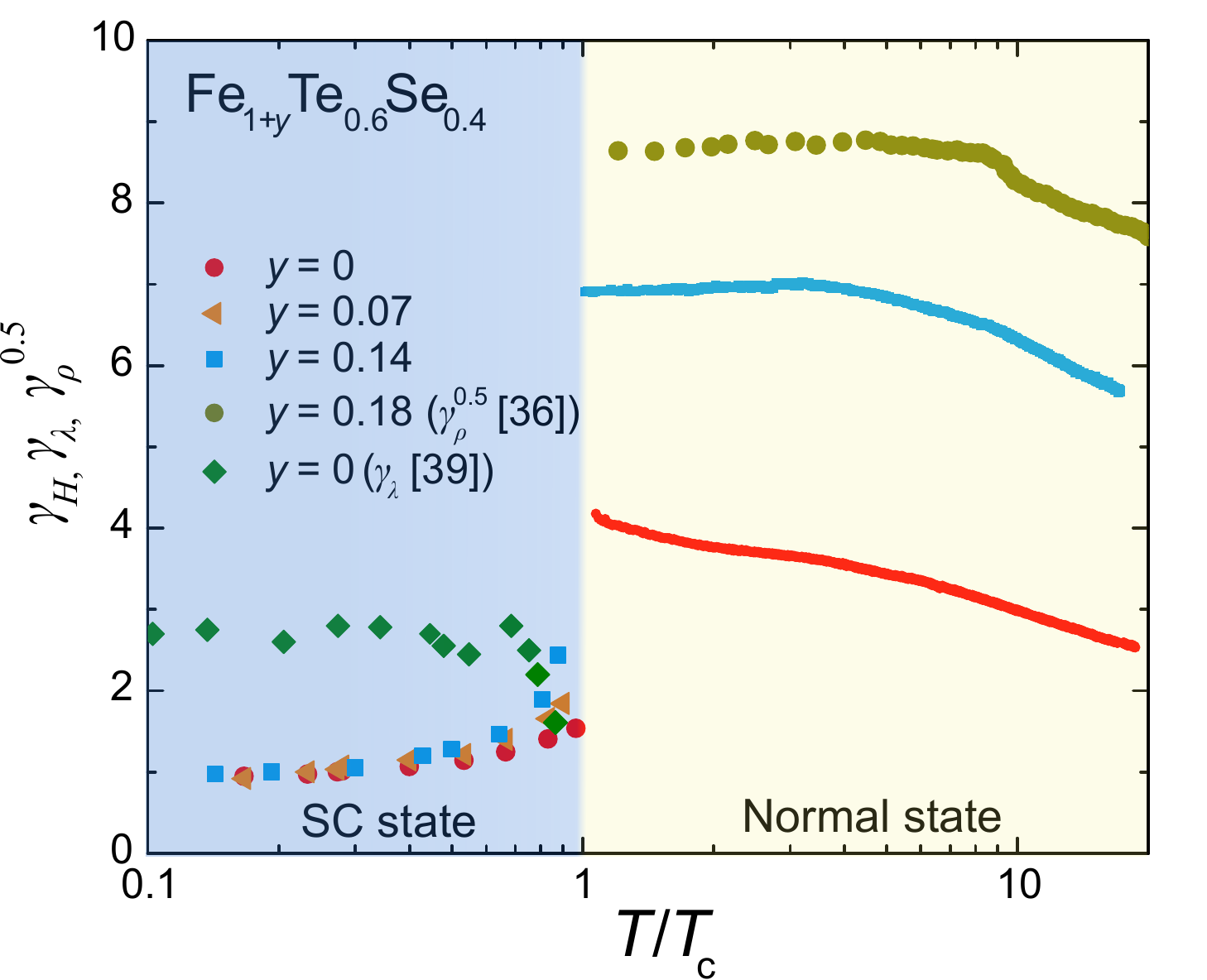}
	\caption{Logarithmic plots of the SC state anisotropy $\gamma_{H}$ and square root of the normal-state anisotropy $\gamma_{\rho}^{1/2}$ as a function of the reduced temperature ($T/T_{\rm{c}}$), at the range of 1.5 K $\leq$ $T$ $\leq$ 300 K. For comparison, normal-state anisotropy of Fe$_{1.18}$Te$_{0.6}$Se$_{0.4}$ from Ref. \cite{LiuPhysRevB.80.174509} and SC state anisotropy obtained from the penetration-depth measurements $\gamma_{\lambda}$ $\equiv$ $\lambda_c$/$\lambda_{ab}$ of Fe$_{1.0}$Te$_{0.6}$Se$_{0.4}$ \cite{BendelePhysRevB_lambda} are also plotted.}\label{}
\end{figure}

Obviously, the anisotropy in the normal state is much larger than that in the SC state. For Fe$_{1.0}$Te$_{0.6}$Se$_{0.4}$, $\gamma_{\rho}^{0.5}$ resides in the region of 2.5 - 4. However, it increases up to $\sim$ 5.7 - 7 in Fe$_{1.14}$Te$_{0.6}$Se$_{0.4}$, and $\sim$ 7.7 - 8.8 in Fe$_{1.18}$Te$_{0.6}$Se$_{0.4}$ \cite{LiuPhysRevB.80.174509}. The values of $\gamma_H$ and $\gamma_{\rho}$ are also summarized in Table \Romannum{1}. In order to resolve the observed discrepancy between the SC and normal-state anisotropies, we need to reconsider the empirical relation of $\gamma_{H}$ $\sim$ $\gamma_{\rho}^{0.5}$. According to the Drude Model, $\gamma_{\rho}$ can be expressed as
\begin{equation}
\label{eq.3}
\gamma_{\rho}=\frac{\rho_c}{\rho_{ab}}=\frac{m^*_c}{ne^2\tau_c}/\frac{m^*_{ab}}{ne^2\tau_{ab}}=\frac{\tau_{ab}}{\tau_c}\frac{m^*_c}{m^*_{ab}},
\end{equation}
where $n$ is the charge carrier density, and $\tau_{ab}$ and $\tau_c$ are the carrier scattering times in the $ab$-plane and along the $c$ axis, respectively. The empirical relation of $\gamma_{H}$ $\sim$ $\gamma_{\rho}^{0.5}$ is approximately obtained by assuming the isotropic scattering. Therefore, the different anisotropy between the SC and normal state, observed universally for samples with different amounts of excess Fe, clearly shows the contribution of the anisotropic scattering time $\tau$. By assuming that the ratio of $m^*_c$/$m^*_{ab}$ is continuously connected at $T_{\rm{c}}$, we roughly estimate the ratio of $\tau_{ab}$/$\tau_c$ (=$\gamma_{\rho}$/$\gamma_{H}^2$) as $\sim$ 7.77 for Fe$_{1.0}$Te$_{0.6}$Se$_{0.4}$ and $\sim$ 7.80 for Fe$_{1.14}$Te$_{0.6}$Se$_{0.4}$. Therefore, the large discrepancy between the SC and normal-state anisotropies is due to the anisotropy of the scattering. 

Besides, the anisotropy of $\tau_{ab}$/$\tau_c$ is almost identical for crystals with different amounts of excess Fe, which indicates that the scattering from excess Fe should be isotropic. The excess Fe in the interstitial position is reported to be strongly magnetic, which provides local moments that interact with the Fe in the FeTe/Se plane \cite{ZhangLijunPRB}. Neutron scattering measurements find out that the excess Fe in Fe$_{1+y}$Te$_{1-x}$Se$_x$ will cause spin clusters involving more than 50 Fe in the nearest two neighboring Fe-layers \cite{ThampyPRL}. Considering the amount of excess Fe is as large as 14\% in Fe$_{1.14}$Te$_{0.6}$Se$_{0.4}$, the influence of such magnetic clusters to the scattering should be more extensive, compared to the case of isolated impurities. Our observation of the isotropic scattering from excess Fe suggests that the magnetic moment should be randomly orientated without order.

\section{conclusions}

We investigated the reported discrepancy between the SC and normal state anisotropies of Fe$_{1+y}$Te$_{1-x}$Se$_{x}$ superconductors by probing the anisotropies of crystals with controlled amounts of excess Fe. The SC-state anisotropy $\gamma_{H}$ is found to be in the range of 1 $\sim$ 2.5 in the crystals with excess Fe ranging from 0 to 14\%, while the normal-state anisotropy $\gamma_{\rho}$ shows a much larger value of 17 $\sim$ 50 at the temperature above $T_{\rm{c}}$. Combining the results of $\gamma_{H}$ and $\gamma_{\rho}$, we found out that such discrepancy originates from a large anisotropic scattering time $\tau_{ab}$/$\tau_c$ $\sim$ 7.8 in the normal state. Besides, the $\tau_{ab}$/$\tau_c$ is found to be independent of the excess Fe.

\acknowledgements
The authors would like to thank Dr. Shin-ya Ayukawa and Mr. Daiki Kakehi for their stimulating pioneer work, and Dr. Jinsheng Wen from Nanjing University, and Dr. Peng Zhang from ISSP, the University of Tokyo for the helpful discussions. The present work was partly supported by the National Key R\&D Program of China (Grant No. 2018YFA0704300), and KAKENHI (JP20H05164, 19K14661, 18K03547, 16K13841, and 17H01141) from JSPS. FIB microfabrication performed in this work was supported by Center for Instrumental Analysis, College of Science and Engineering, Aoyama Gakuin University. ICP analyses were performed at Chemical Analysis Section in Materials Analysis Station of NIMS.

S.Y. and Y.P. contributed equally to this paper.


\bibliography{references}

\end{document}